\documentclass[12pt]{article}

\pdfoutput=1
\usepackage{graphicx} 
\usepackage{xspace}
\usepackage{url}
\usepackage{comment}
\usepackage{amsmath}
\usepackage{amssymb}
\usepackage{fancyhdr}  
\usepackage{hepunits}
\usepackage[usenames,dvipsnames]{color}  
\usepackage{enumitem}
\usepackage{multirow}
\usepackage{lineno}
\usepackage{pdfpages}

\usepackage[a4paper, total={17cm, 24.cm}, headheight=0.6cm]{geometry}
\usepackage{subcaption}  
\usepackage[pdftex,bookmarks,hidelinks]{hyperref}

\usepackage[normalem]{ulem}
\def\phaseone{Phase~I\xspace} 
\def\phasetwo{Phase~II\xspace} 

\def\expshort{DUNE\xspace}
\def\dune{\expshort}

\newcommand{\rms}{RMS\xspace} 

\newcommand{\deltacp}{\ensuremath{\delta_{\rm CP}}\xspace}   

\def\argon40{${}^{40}$Ar}       
\def\Ar39{$^{39}$Ar}
\def\Cl40{$^{40}$Cl}
\def\K40{$^{40}$K}
\def\B8{$^{8}$B}
\newcommand{\lsim}{{\;\raise0.3ex\hbox{$<$\kern-0.75em\raise-1.1ex\hbox{$\sim$}}\;}}
\newcommand{\gsim}{{\;\raise0.3ex\hbox{$>$\kern-0.75em\raise-1.1ex\hbox{$\sim$}}\;}}
\newcommand{\beq}{\begin{equation}}
\newcommand{\eeq}{\end{equation}}
\newcommand{\bea}{\begin{eqnarray}}
\newcommand{\eea}{\end{eqnarray}}

\mathchardef\minus="002D

\DeclareSIUnit \c {$c$}
\DeclareSIUnit\magn{$\times$}
\DeclareSIUnit\min{min}
\DeclareSIUnit\hr{hr}
\DeclareSIUnit\hrs{hrs}
\DeclareSIUnit\week{week}
\DeclareSIUnit\month{mo}
\DeclareSIUnit\months{mos}
\DeclareSIUnit\year{yr}
\DeclareSIUnit\years{years}
\DeclareSIUnit\yr{yr}
\DeclareSIUnit\standard{std}
\DeclareSIUnit\str{sr}
\DeclareSIUnit\ppm{ppm}
\DeclareSIUnit\ppb{ppb}
\DeclareSIUnit\ppt{ppt}
\DeclareSIUnit\pe{PE}
\DeclareSIUnit\spe{SPE}
\DeclareSIUnit\pdm{PDM}
\DeclareSIUnit\ev{events}
\DeclareSIUnit\ct{counts}
\DeclareSIUnit\neutron{\mbox{$n$}}
\DeclareSIUnit\smp{samples}
\DeclareSIUnit\Sample{S}
\DeclareSIUnit\ch{ch}
\DeclareSIUnit\hit{hit}
\DeclareSIUnit\hits{hits}
\DeclareSIUnit\bin{(\mbox{5-PE}~bin)}
\DeclareSIUnit\sgm{\mbox{$\sigma$}}
\DeclareSIUnit\rms{RMS}
\DeclareSIUnit\keVee{\mbox{keV$_{e{\rm e}}$}}
\DeclareSIUnit\keVr{\mbox{keV$_{\rm nr}$}}
\DeclareSIUnit\eVee{\mbox{eV$_{\rm ee}$}}
\DeclareSIUnit\eVr{\mbox{eV$_{\rm nr}$}}
\DeclareSIUnit\ph{photon}
\DeclareSIUnit\el{\mbox{$e^-$}}
\DeclareSIUnit\pm{\mbox{PMT}}
\DeclareSIUnit\pixel{\mbox{pixel}}
\DeclareSIUnit\inch{''}
\DeclareSIUnit\foot{'}
\DeclareSIUnit\bit{bit}
\DeclareSIUnit\sample{samples}
\DeclareSIUnit\barn{barn}
\DeclareSIUnit\bara{bar}
\DeclareSIUnit\bar{bar}
\DeclareSIUnit\barg{barg}
\DeclareSIUnit\mlardepth{\mbox(meter~of~\LAr~depth)}
\DeclareSIUnit\Curie{Ci}
\DeclareSIUnit\PSI{psi}
\DeclareSIUnit\psia{psia}
\DeclareSIUnit\atm{atm}
\DeclareSIUnit\psf{psf}
\DeclareSIUnit\pcf{pcf}
\DeclareSIUnit\parsec{pc}
\DeclareSIUnit\cps{cps}
\DeclareSIUnit\slpm{\SI{}{\liter\per\minute}}
\DeclareSIUnit\rpm{rpm}
\DeclareSIUnit\mwe{\mbox{m.w.e.}}
\DeclareSIUnit\liveday{\mbox{live-days}}
\DeclareSIUnit\days{\mbox{days}}
\DeclareSIUnit\miles{\mbox{miles}}
\DeclareSIUnit\lumens{\mbox{lm}}
\DeclareSIUnit\degreeC{\mbox{$^{\circ}$C}}
\DeclareSIUnit\degreeF{\mbox{$^{\circ}$F}}
\DeclareSIUnit\electron{\mbox{$e^-$}}
\DeclareSIUnit\Euro{\mbox{\euro}}
\DeclareSIUnit\cph{cph}
\DeclareSIUnit\neq{neq}
\DeclareSIUnit\normal{\mbox{N}}
\DeclareSIUnit\USD{\mbox{\$}}
\DeclareSIUnit\Vpercm{\mbox{V/cm}}
\DeclareSIUnit\kV{\mbox{kV}}

\DeclareSIUnit \mm {\milli\meter}
\DeclareSIUnit \cm {\centi\meter}
\DeclareSIUnit \us {\micro\second}
\DeclareSIUnit \ms {\milli\second}
\DeclareSIUnit \pA {\pico\ampere}
\DeclareSIUnit \pC {\pico\coulomb}
\DeclareSIUnit \fC {\femto\coulomb}
\DeclareSIUnit \fF {\femto\farrad}
\DeclareSIUnit \pF {\pico\farrad}
\DeclareSIUnit \mV {\milli\volt}
\DeclareSIUnit \kV {\kilo\volt}
\DeclareSIUnit \V {\volt}
\DeclareSIUnit \GOhm {\giga\ohm}
\DeclareSIUnit \MOhm {\mega\ohm}
\DeclareSIUnit \ton {\tonne}
\DeclareSIUnit \kton {\kilo\tonne}
\DeclareSIUnit \kt {\kilo\tonne}
\DeclareSIUnit \Mt {\mega\tonne}
\DeclareSIUnit \eV {\electronvolt}
\DeclareSIUnit \keV {\kilo\electronvolt}
\DeclareSIUnit \MeV {\mega\electronvolt}
\DeclareSIUnit \GeV {\giga\electronvolt}
\DeclareSIUnit \km {\kilo\meter}
\DeclareSIUnit \kW {\kilo\watt}
\DeclareSIUnit \MW {\mega\watt}
\DeclareSIUnit \MHz {\mega\hertz}
\DeclareSIUnit \kHz {\kilo\hertz}
\DeclareSIUnit \mrad {\milli\radian}
\DeclareSIUnit \year {year}
\DeclareSIUnit \POT {POT}
\DeclareSIUnit \sig {$\sigma$}
\DeclareSIUnit\parsec{pc}
\DeclareSIUnit\lightyear{ly}
\DeclareSIUnit\foot{ft}
\DeclareSIUnit\ft{ft}

\title{%
The DUNE \phasetwo Detectors  \\ \bigskip
\large Input to the European Strategy for Particle Physics - 2026 Update }
\author{The DUNE Collaboration
\footnote{Contact persons: Sergio Bertolucci (Sergio.Bertolucci@cern.ch), Sowjanya Gollapinni (sowjanya@lanl.gov)}
}

\date{\today}

\begin{document}

\maketitle

\begin{abstract}
    The international collaboration designing and constructing the Deep Underground Neutrino Experiment (DUNE) at the Long-Baseline Neutrino Facility (LBNF) has developed a two-phase strategy for the implementation of this leading-edge, large-scale science project. The 2023 report of the US Particle Physics Project Prioritization Panel (P5) reaffirmed this vision and strongly endorsed DUNE Phase~I and Phase~II, as did the previous European Strategy for Particle Physics. The construction of DUNE Phase~I is well underway. DUNE Phase II consists of a third and fourth far detector module, an upgraded near detector complex, and an enhanced $>2$\,MW beam. The fourth FD module is conceived as a ``Module of Opportunity'', aimed at supporting the core DUNE science program while also expanding the physics opportunities with more advanced technologies. 

    The DUNE collaboration is submitting four main contributions to the 2026 Update of the European Strategy for Particle Physics process. This submission to the “Detector instrumentation” stream focuses on technologies and R\&D for the DUNE \phasetwo detectors. Additional inputs related to the DUNE science program, DUNE software and computing, and European contributions to Fermilab accelerator upgrades and facilities for the DUNE experiment, are also being submitted to other streams.

\end{abstract}

\thispagestyle{empty} 

\newpage
\pagenumbering{arabic}

\section{Scientific Context}
\label{sec:context}
\addcontentsline{toc}{section}{Executive summary}

The preponderance of matter over antimatter in the early universe, the dynamics of the supernova neutrino bursts (SNBs) that produced the heavy elements necessary for life, and the nature of physics beyond the Standard Model (BSM) are mysteries at the forefront of particle physics and astrophysics, and key to understanding the evolution of our universe.

The Deep Underground Neutrino Experiment (DUNE) will address these questions in a multidecadal science program with its world-leading liquid argon (LAr) detector technology. The international DUNE collaboration, hosted by the Fermi National Accelerator Laboratory (FNAL), is designing, developing, and constructing a near detector (ND) complex at FNAL (the near site) and a suite of four large detector modules 1300\,km downstream at the Sanford
Underground Research Facility (SURF) in South Dakota (the far site). These detectors will record neutrinos over a wide energy range, originating from a new high-intensity neutrino beamline at FNAL. The modular far detector (FD) will also detect neutrinos produced in the atmosphere and from astrophysical sources. The beamline as well as the excavations, infrastructure, and facilities for housing and supporting the DUNE detectors are provided by the Long-Baseline Neutrino Facility (LBNF).

The DUNE Collaboration was launched in 2015, following the recommendations of the 2013 update of the European Strategy for Particle Physics~\cite{2013europeanstrategy} and of the 2014 Report of the US Particle Physics Project Prioritization Panel (P5)~\cite{2014p5report}. DUNE and LBNF will complete this project in two phases,  as summarized in Table~\ref{tab:phases}, based on the availability of resources and the ability to reach science milestones. The latest P5 report released in December 2023 reaffirmed this vision~\cite{2023p5report}. 

The \phaseone beamline will produce a wideband neutrino beam with up to $1.2$~MW beam power, designed to be upgradable to $>2$\,MW. The \phaseone ND includes a moveable LArTPC with pixel readout (ND-LAr), integrated with a downstream muon spectrometer (TMS)~\cite{DUNE:2021tad}, and an on-axis magnetized neutrino detector (SAND) further downstream. The ND-LAr+TMS detector system can be moved sideways over a range of off-axis angles and neutrino energies (the DUNE-PRISM concept), for an optimal characterization of the neutrino-argon interactions. The \phaseone FD includes two LArTPC modules, each containing 17\,kt of LAr. The far detector module 1 (FD1) is a horizontal drift time projection chamber (TPC), as developed and operated in ProtoDUNE at CERN~\cite{DUNE:2020txw}. The far detector module 2 (FD2) is a vertical drift TPC~\cite{DUNE:2023nqi}. For the cryogenic infrastructure in support of the two LArTPC modules, \phaseone will include two large cryostats (one per FD module), 35\,kt of LAr, and three nitrogen refrigeration units.

\begin{table}[bp]
    \centering
    \begin{tabular}{|p{3.3cm}|p{4.7cm}|p{4.7cm}|p{2.3cm}|} \hline
        Parameter  & \phaseone      & \phasetwo & Impact \\ \hline
        FD mass    & 2 FD modules ($>20$\,kt fiducial) & 4 FD modules ($>40$\,kt fiducial LAr equivalent) & FD statistics \\  \hline
        Beam power & 1.2\,MW & $>2$\,MW   & FD statistics \\ \hline
        ND configuration  & ND-LAr+TMS, SAND & ND-LAr, ND-GAr, SAND   & Systematics \\ \hline
    \end{tabular}
    \caption{A high-level description of the two-phased approach to DUNE. The ND-LAr detector, including its capability to move sideways (DUNE-PRISM), and SAND are present in both phases of the ND. Note that the non-argon options currently under consideration for \phasetwo near and far detectors are not shown.}
    \label{tab:phases}
\end{table}

The construction of the first project phase (\phaseone), funded through commitments by a coalition of international funding agencies, is well underway. Its successful completion is currently the collaboration's highest priority. Excavation at the far site is complete, and fabrication of various beamline and detector components for \phaseone is progressing well. The facilities currently being constructed by LBNF at both the near and far sites are designed to host the full scope (\phaseone and \phasetwo) of DUNE.

\phasetwo of DUNE~\cite{DUNE:2024wvj} encompasses an enhanced multi-megawatt beam, the third and fourth FD modules, and an upgraded ND complex. The primary objective of \dune \phasetwo is a set of precise measurements of the parameters of the neutrino mixing matrix, $\theta_{23}$, $\theta_{13}$, $\Delta m^{2}_{32}$, and \deltacp, to establish Charge Conjugation-Parity Symmetry Violation (CPV) over a broad range of possible values of \deltacp, and to search for new physics in neutrino oscillations. DUNE also seeks to detect low-energy neutrinos from astrophysical sources. The additional mass brought by the \phasetwo FD modules will increase the statistics of a supernova burst signal and extend DUNE's reach beyond the Milky Way. The \phasetwo design concepts could also enable sensitive searches for new physics with solar neutrinos by lowering the detection threshold and by reducing background rates. Finally, \phasetwo will expand DUNE's new physics discovery reach for rare processes at the ND and FD sites, and for non-standard neutrino oscillations.

The \phasetwo R\&D program is a global effort with contributions from all DUNE partners and potential new collaborators.  Part of the R\&D described in this document is carried out within the framework of the ECFA detector R\&D collaborations and those being formed under the umbrella of the Coordinating Panel for Advanced Detectors (CPAD) in the US. 

Support by the European science community is critical to the success of DUNE's ground-breaking neutrino science program. A $39\%$ fraction of DUNE collaborators are based at European universities and research institutions. They provide key contributions to the Far and Near detector components, computing, as well as to the LBNF beam. These contributions, funded largely by European national funding agencies, 
are key to the design and construction of LBNF and DUNE Phase I. The R\&D and validation of the detector technologies for DUNE's first two far detector modules has been enabled by the Neutrino Platform at CERN through its ProtoDUNE program, which has been an extraordinary success. 
Finally, CERN provides key LBNF infrastructure, in particular through the
procurement of the FD cryostats. CERN also plays an
important role as European hub for the DUNE Collaboration, regularly hosting collaboration meetings
and workshops. 

DUNE relies on CERN's continued support, specifically by:

\begin{itemize}
    \item providing infrastructure, technical expertise, resources, support for DUNE R\&D, and detector technology and physics validation through the Neutrino Platform and the ProtoDUNE facilities;
    \item additional support of the far site facility (LBNF) through the procurement of cryostats for the third and fourth modules;
    \item continued technical and engineering support during the DUNE construction phase;
    \item CERN support for addressing growing DUNE computing needs;
    \item support and coordination of the DRD collaborations;
    \item providing important physics and technical leadership on DUNE through CERN's EP-NU group and CERN’s Neutrino Platform;
    \item providing facilities for GeV-scale beam tests for detector R\&D at CERN, and by supporting dedicated experiments to measure hadron production cross-sections for the DUNE environment.
\end{itemize}

\section{Objectives}
\label{sec:objectives}

\subsection{Key performance indicators for \phasetwo Far Detector Modules}
\label{subsec:obj_fd}

The overall goal of the \phasetwo FD modules, FD3 and FD4, is to support the core DUNE physics program while also expanding the physics opportunities with more advanced technologies.

For long-baseline physics, FD3 and FD4 will provide the additional exposure and improved statistical precision that is critical to achieve the full \deltacp sensitivity. They also offer the opportunity to improve the neutrino energy reconstruction and the neutrino interaction classification through optimized charge and photon readout systems. It is crucial that the data from these modules be combined with data from FD1 and FD2, with the systematic constraints from the ND applied to all FD modules. For this reason, the most straightforward approach is for FD3 and FD4 to be LAr TPCs, so that DUNE would immediately benefit from the $\nu$-Ar measurement program of the ND. 

In the case of neutrino astrophysics and other low-energy physics opportunities, the potential to enhance the physics scope of DUNE with lower energy thresholds has been attracting significant attention~\cite{Andringa:2023aax}. Large LAr TPCs have demonstrated charge detection thresholds well below 1~MeV. However, the ultimate analysis thresholds will be set by radiological backgrounds and by the photon detection coverage. In \phaseone, for steady state sources such as solar neutrinos, analysis thresholds of about 10~MeV will be reached. For transient sources such as supernova neutrino bursts, lower analysis thresholds are possible. In \phasetwo, with greater control of radioactive backgrounds from neutrons, gammas and radon, and better energy resolution, we aim to lower the analysis thresholds down to approximately 5~MeV for all astrophysical neutrino sources. With such a threshold, an extended SNB neutrino program can be envisaged, with improved reach in terms of supernova distance sensitivity (to the Magellanic Clouds), for elastic scatters with improved directionality, and to the (softer energy) early or late parts of the supernova neutrino flux. A low-energy threshold could also allow a precision solar neutrino program to explore solar-reactor oscillation tensions and non-standard interactions. 

Key performance indicators for LAr-based \phasetwo FD modules’ technologies are: signal-to-noise ratio achieved by charge readouts, power consumption of native 3D charge readouts; spatial resolution; electron lifetime; overall light yield and uniformity of response of photon detector systems; mitigation of external backgrounds by shield systems.


\subsection{Key performance indicators for the \phasetwo Near Detector}
\label{subsec:obj_nd}

In \phasetwo, DUNE will have accumulated FD statistics of several thousand oscillated electron neutrinos, resulting in statistical uncertainties at the few-percent level on the number of electron appearance events. To reach the physics goals of DUNE, a similar level of systematic uncertainty must be achieved, which requires precise constraints from the ND. The \phasetwo ND should maintain argon as the primary target nucleus. It should also perform better than the \phaseone ND on the following key performance indicators: particle identification (PID) as a function of track momenta and angles, tracking thresholds achievable for protons and pions, mitigation of secondary interactions in the tracker volume, angular acceptance as a function of track momenta, charge sign identification as a function of track momentum.

Employing an argon target will ensure that constraints from the \phasetwo ND can be applied directly to the argon-based FD without any extrapolation in atomic number. A broad acceptance and high PID efficiency will enable exclusive final states to be identified, which will improve the constraints on neutrino interaction modeling. Low thresholds will make the \phasetwo ND highly sensitive to nuclear effects. Magnetization will ensure sign selection at all energies and angles, for both charged leptons and charged pions. This will be necessary to separate neutrino from antineutrino interactions and for fully identifying exclusive final states.

Beyond long-baseline physics, the high intensity and high energy of the LBNF proton beam enables DUNE to search for a wide variety of long-lived, exotic particles that are produced in the target and decay in the ND. A \phasetwo ND system with lower density and improved rejection of neutrino interaction backgrounds expands the reach for these searches.
\section{Methodology}
\label{sec:methodology}

The DUNE \phasetwo white paper provides a significant amount of details on the \phasetwo detector options at both the far (SURF) and near (FNAL) sites, please see \cite{DUNE:2024wvj} and references therein. Here, we summarize the present thinking on the reference detector designs for each, as of the writing of this document. Full detector solutions will be defined in the forthcoming years through dedicated design reports as prototyping matures and target technologies are fully demonstrated.

The LBNF/DUNE project performed an Environmental Assessment evaluation of its potential environmental impacts and the safety and health hazards during construction and operation of the project, and a "Finding of No Significant Impact" was issued~\cite{edms-2808692}.

\subsection{The \phasetwo Far Detector Modules}
\label{subsec:methodology_fd}

The DUNE FD2 vertical drift technology forms the basis for the reference design for FD3 and FD4. As such, the R\&D for FD3 and FD4 is primarily focused on upgraded photon detection and charge readout systems for the vertical drift layout. A non-LAr option such as hybrid Cherenkov and scintillator detector, for example using water-based liquid scintillator, is also under consideration as an alternative technology for FD4. A summary of technologies under consideration and a road map towards defining complete detector designs is described below.

\subsubsection{Technologies Under Consideration}
\paragraph{Optimized photon readout systems (APEX, PoWER).} 
The \textbf{APEX} concept (Aluminum profiles with embedded X-ARAPUCA) integrates a large-area photon detection system into the detector module’s field cage~
\cite{PhysRevD.111.032007,Shi:2025rob,Marinho:2025cyc}.
This solution is derived from the well-established technology developed for FD2. APEX can provide up to about 60\% coverage of the surface enclosing the LArTPC active volume if the four field cage vertical walls are fully instrumented. \textbf{PoWER} (Polymer Wavelength shifter and Enhanced Reflection) is another recent proposal~\cite{Steklain:2025flt} for a non-anode-based improved photon detector system. It involves full coverage of the field cage with polymeric wavelength shifting foils, light detection units based on large arrays of SiPMs mounted on the cryostat membrane, and full coverage of the membrane with reflector panels.
 
\paragraph{Strip-based charge readout (CRP).}
The PCB-based vertical drift anodes, called CRPs, are made up of two stacked PCBs, providing three projective views. CRPs have been successfully demonstrated at full scale, have been installed in ProtoDUNE-VD in the NP02 cryostat at CERN, and will be deployed in the FD2 cryostat at SURF. 
 
\paragraph{Pixel-based charge readout (LArPix, Q-Pix, GAMPix).}
A pixel-based readout could replace the multi-layer strip-based readout with a single-layer grid of charge-sensitive pixels at mm-scale granularity. Instrumenting each pixel with a dedicated electronics channel would achieve a LAr TPC with true and unambiguous 3D readout. Given channel densities of $\mathcal{O}$(10$^5$) pixels per m$^2$ of anode, pixel readout would require operation at $\mathcal{O}$(100)~$\mu$W power consumption per channel. \textbf{LArPix} is a complete pixel readout system for LAr TPCs. It has been developed as the baseline technology of \phaseone ND-LAr, and already meets most of the requirements for deployment in a future FD module. The system relies on the LArPix ASIC, a 64-channel detector system-on-a-chip that includes analog amplification, self-triggering, digitization, digital multiplexing, and a configuration controller. \textbf{Q-Pix} is a novel pixel-based technology for low-threshold, high-granularity, low-throughput readout. It is ideally suited to the low data rate readout environment of the DUNE FD modules. Another technology that has recently been proposed~\cite{Shutt:2024che} is \textbf{GAMPix} (Grid-Activated Multi-scale Pixels), combining an external induction grid with a pulsed-power pixel plane. This arrangement may yield significantly lower electronic noise and energy thresholds, as well as provide a charge-only drift length measurement, while maintaining low power consumption and data rates.

\paragraph{Optical-based charge readout (ARIADNE).}
In the ARIADNE technology, the secondary scintillation light produced in amplification structures within the gaseous argon atmosphere above the LAr surface can be captured by fast cameras to reconstruct the primary ionization track in 3D. The optical-based readout shares the same physics benefit as the pixel-based charge readout solutions in providing a native 3D readout, with demonstrated mm-scale spatial resolution. The data-driven readout with native zero suppression yields a very efficient raw data storage. The overall optical gain and the low-noise readout environment enable low-threshold operation. 

\paragraph{Integrated charge and light readout on anode (SoLAr, LightPix, Q-Pix-LILAr).}
DUNE is also pursuing the integration of both light and charge detection modes on the anode into a single detector element. This would offer intrinsic fine-grained information for both charge and light, an enhancement in the amount of light collected near the anode, and a simplification in the detector design. The \textbf{SoLAr} readout unit is a pixel tile based on PCB technology that embeds charge readout pads to collect drifting charges, and highly efficient VUV SiPMs to collect photons. It aims to achieve a low energy threshold with excellent energy resolution and background rejection through pulse-shape discrimination. \textbf{LightPix} is a variant of the LArPix ASIC that has been designed for scalable readout of very large arrays of SiPMs and that could be used as a readout unit for an anode-based light pixel solution. The \textbf{Q-Pix-LILAr} concept relies on coating a charge readout pixel with a type of photo-conductive material that, when struck by a VUV photon, would generate a signal that could be detected by the same readout scheme used for the ionization charge. 

\paragraph{Water-based, hybrid Cherenkov plus scintillation, detection concept (THEIA).}
This concept would provide separate Cherenkov and scintillation light detection by the use of novel liquid scintillator, fast timing, and spectral sorting technology. It would enable particle identification, sensitivity to particle direction, excellent energy and vertex resolution, and sub-Cherenkov-threshold particle detection. 

\subsubsection{Road map Towards Complete Detector Designs}
The technologies summarized above will form the building blocks to define complete detector designs for FD3 and FD4. These are not standalone, and most of them can be combined or integrated together. Most of these candidate systems are either further developments of the current systems or replacements based on technologies that are already under active R\&D or in early prototyping phases, see Sec.~\ref{sec:readiness}. 

The reference design for FD3 is a vertical drift LAr TPC similar to FD2 with only modest upgrades planned to the Charge Readout Planes (CRPs) and the photon detection system along with improvements based on lessons learned from FD2. Continued R\&D beyond the current CRP design for FD2 will focus on optimizations to reduce cost or to improve performance, by optimizing strip pitch, length, and orientation, as well as on streamlining CRP construction techniques. Upgrades to the FD2 charge readout electronics are also possible. 

The FD3 photon detection system would be composed of X-ARAPUCA-based photon detector modules read by SiPMs using Power-over-Fiber (PoF) and Signal-over-Fiber (SoF), similar to FD2. The detailed design of the FD3 photon detection system (location and design of PD modules, optical coverage, readout electronics solution) will evolve from the FD2 design to incorporate improvements from APEX R\&D. Increased photon coverage with APEX improves the energy resolution obtained from scintillation light, as well as the triggering using the light signal, enhancing the low-energy physics reach. We envisage LAr doping as for FD2, via the addition of trace (ppm-level) amounts of liquid xenon, to decrease the amount of Rayleigh scattering and to narrow the timing distribution of the scintillation light.

The current concepts for FD4 introduce further improvements beyond what is planned for FD3. The reference design for FD4 is a vertical drift LAr TPC with a central cathode and two anodes with pixel-based readouts. The projective readout of CRPs would be replaced by a native 3D charge readout system, either employing charge pixels (LArPix or Q-Pix) or through an optical-based charge readout (ARIADNE). The anode pixels may also serve as scintillation light detection units (SoLAr, LightPix and Q-Pix-LiLAr). Native 3D charge readout resolves ambiguities that can arise with projective readout when two or more particles overlap in a projection, or when a particle track points along the strip direction. This is especially beneficial in reconstructing higher-multiplicity GeV-scale neutrino interactions. The symmetric TPC configuration may in principle allow for implementation of different pixel-based solutions at the top and bottom anodes, depending on the R\&D outcome and available resources. 

A single-drift LAr TPC solution for FD4 with a unique ARIADNE-based anode plane on top and the cathode placed at the bottom of the detector is also possible. This solution would require upgrades to the high voltage system to accommodate a longer (13~m) drift and full-scale prototype tests with commercial 600~kV power supplies. 

In DUNE \phaseone, dominant backgrounds to MeV-scale neutrinos will be produced by neutrons and gammas of radiogenic origin from outside the detector (SURF cavern rock and shotcrete). When neutrons are captured in the LAr, they can produce gamma cascades which Compton scatter or pair produce electrons that directly mimic the charged-current neutrino signals. Mitigation of these external neutrons is essential for the argon-based readout technologies. The design of compact shields~\cite{Capozzi:2018dat} that can fit in the limited DUNE cavern space, or within the cryostat structure, will be necessary. A layer of 30--40\,cm of water, polyethylene or borated polyethylene, is sufficient to attenuate the neutron flux by 3 orders of magnitude. 

A water-based, hybrid scintillation and Cherenkov detector would provide a different technology for FD4 compared to FD1, FD2, and FD3, and currently forms the basis of the alternative FD4 concept. This module could substantially reduce backgrounds at low energy compared to LAr, extending the solar neutrino reach down to CNO energies and adding sensitivity to other low-energy signals like the diffuse supernova background. Such a detector could also contribute to the long-baseline oscillation program, provided systematic uncertainties could be constrained at a similar level as in LAr, for example by including a dedicated water-based liquid scintillator near detector.

The cryogenic infrastructure at the far site will have to be upgraded for \phasetwo to provide refrigeration for up to an additional 35 kt of LAr, and the cryostats for the additional LAr-based FD modules. It is assumed that the cost sharing between US and non-US partners in DUNE for the full \phasetwo FD scope (detector components, detector installation, cryostats, cryogenic system upgrade, and cryogens) will mirror the one currently in place for the \phaseone FD scope. CERN support for the cryostats of the \phasetwo FD modules, as well as for the validation at large scale of the proposed FD technologies, is essential.


\subsection{The \phasetwo Near Detector}
\label{subsec:methodology_nd}

For \phasetwo, an improved tracker concept called ND-GAr would replace the Muon Spectrometer (TMS) downstream of ND-LAr. The baseline ND-GAr concept is based on a central high-pressure gaseous argon time projection chamber (HPgTPC). The HPgTPC will be surrounded by a calorimeter, with both situated in a 0.5~T magnetic field generated by superconducting coils. A muon-tagging system is integrated with the magnet return yoke. A photon detection system (PDS) may also prove necessary to reduce pileup and to provide the event $t_{0}$ in events that do not reach the calorimeter.

A cylindrical volume with a diameter and length both of roughly 5\,m, and gas at 10\,bar, would have a fiducial mass of nearly one ton of argon, yielding approximately one million neutrino interactions per year. The entire ND-GAr system will move perpendicularly to the beam direction together with ND-LAr, as part of the DUNE-PRISM concept.

In addition to ND-GAr, upgrades to the \phaseone near detectors, ND-LAr and SAND, are also possible. In the case FD4 uses a non-argon-based technology, several near detector options are under consideration to match the far detector target and technology at the near site. This includes adding additional nuclear targets in SAND, integrating water-based liquid scintillator targets in the ND-GAr calorimeter, or introducing a new, dedicated detector.


\section{Readiness and R\&D Goals}
\label{sec:readiness}

\begin{table}[tbp]
\centering
\begin{tabular}{|p{2.5cm}|p{6.5cm}|p{6.5cm}|} \hline
\textbf{Technology} & \textbf{Prototyping Plans} & \textbf{Key R\&D Goals} \\ \hline 

\textbf{CRP} & 
1. Cold Box tests at CERN. \newline
2. ProtoDUNE at CERN. 
& Port LArASIC to 65~nm process. \\ \hline

\textbf{APEX} & 
\multirow{2}{7.5cm}{1. 50\,L \& 2-ton prototypes at CERN. \newline
2. $\mathcal{O}$(10)-channel demonstrator at FNAL. \newline
3. ProtoDUNE at CERN.
}
& Mechanical integration of APEX photon detector in field cage. \\ \cline{3-3}
 & & Signal conditioning, digitization and multiplexing in cold. \\ \hline

\textbf{LArPix}, \textbf{LightPix} &
\multirow{3}{7.5cm}{1. 2x2 ND demonstrator at FNAL. \newline
2. Cold Box tests at CERN. \newline
3. ProtoDUNE at CERN.
}
& Micropower, cryo-compatible, detector-on-a-chip ASIC. \\ \cline{3-3}
 & & Scalable integrated 3D pixel anode tile. \\ \cline{3-3}
 & & Digital aggregator ASIC and PCB. \\ \hline

\textbf{Q-Pix}, \textbf{Q-Pix-LILAr} &
\multirow{3}{7.5cm}{1. Prototype chips in small-scale demonstrator. \newline
2. 16 channels/chip prototypes in ton-scale demonstrator at ORNL. \newline 
3. Full 32-64 channel ``physics chip''.
} 
& Charge replenishment and measurement of reset time. \\ \cline{3-3}
 & & Power consumption. \\ \cline{3-3}
 & & R\&D on amorphous selenium-based devices and other photoconductors. \\ \hline

\textbf{ARIADNE} &
\multirow{3}{7.5cm}{1. Glass THGEM production at Liverpool. \newline 
2. ProtoDUNE at CERN.
} 
& Custom optics for TPX3 camera. \\ \cline{3-3}
 & & Light Readout Plane design with glass-THGEMs. \\ \cline{3-3}
 & & Characterization of next-generation TPX4 camera. \\ \hline

\textbf{SoLAr} &
\multirow{2}{7.5cm}{1. Small-size prototypes at Bern. \newline 
2. Mid-scale demonstrator at Boulby.
} 
& Development of VUV-sensitive SiPMs. \\ \cline{3-3}
 & &  ASIC-based readout electronics. \\ \hline

\textbf{Hybrid Cherenkov+ scintillation} &  
\multirow{3}{7.5cm}{1. Prototypes at BNL (1- \& 30-ton), LBNL (\textsc{eos}), FNAL (ANNIE). \newline 
2. BUTTON at Boulby.
} 
& \textsc{theia} organic component manufacturing. \\ \cline{3-3}
 & & \textsc{theia} {\it in situ} purification. \\ \cline{3-3}
 & & Spectral photon sorting (dichroicons). \\ \hline
\end{tabular}
\caption{Prototyping plans and key R\&D goals for the main \phasetwo FD technologies under consideration.} 
\label{tab:FDRandDSummary}
\end{table}

Key challenges related to the technologies under consideration for the \phasetwo FD modules are summarized in Table~\ref{tab:FDRandDSummary}. The table specifies the critical R\&D elements that remain to be addressed and the prototyping phases to be realized before \phasetwo detector technical designs adopting a specific technology can be finalized. All the R\&D activities listed in Table~\ref{tab:FDRandDSummary} are at an advanced stage with a Technology Readiness Level (TRL) of 3 or above\footnote{We adopt the TRL definition as per \cite{trl}.}.

As shown by the prototyping plans in Table~\ref{tab:FDRandDSummary}, the ProtoDUNE detectors at CERN will continue to serve as an important platform to demonstrate several of these FD technologies and their potential for integration into a LAr TPC. The CERN Neutrino Platform is a unique infrastructure worldwide, in this regard. With the ProtoDUNE-II run expected to end in 2025, the DUNE collaboration and the CERN Neutrino Platform leadership have started to plan for future ProtoDUNE runs. Ensuring support and continued operation of the ProtoDUNE facility at CERN will be essential for DUNE, with the aim to test a broad range of novel readouts and inform the technology down-selection for the FD3 and FD4 modules.

The necessary R\&D for the DUNE \phasetwo FD modules, both LAr TPCs and liquid scintillator technologies, are fully integrated within the scope of the CERN DRD2 collaboration~\cite{Detector:2784893}. All DRD2 work packages are directly relevant. The DRD2 organization provides a forum to exploit the technological synergies between the accelerator-based neutrino physics community where DUNE belongs, and the dark matter direct detection and neutrinoless double beta decay communities. It also provides a structure to define the global R\&D program necessary for DUNE \phasetwo, in coordination with the CPAD RDC umbrella in the US.  

For ND-GAr, the critical R\&D item is the testing of the full TPC readout chain from amplification technology to readout electronics at high pressure. Ongoing R\&D thrusts are amplification and electronics. While TPC amplification focused initially on the ALICE-style MWPCs, the current focus is on MPGDs such as GEMs. For the TPC readout, full slice tests of the SAMPA-based electronics have established a clear path to deliver a cost-effective system. An additional beam test is necessary to integrate GEMs with SAMPAs. The focus would be on low-energy beams that match the DUNE requirements. The CERN Neutrino Platform is a potential site for this. A CERN test beam program could also include neutron beams at n\_TOF. Additional ND-GAr R\&D items are high critical temperature superconducting cables for the magnet, and efficient and low-noise readout for primary scintillation light. 

Many of the ND-GAr R\&D themes are being pursued in the context of the CERN DRD1 collaboration on gaseous detectors~\cite{Detector:2784893}. As a large community of 170 institutions and 700 members, with a large experience in gaseous detector development across many different applications, DRD1 could provide needed additional R\&D resources toward realizing ND-GAr.

The ND-LAr and SAND components of the ND may also be upgraded for \phasetwo. A decision on these possible upgrade paths will come after the \phaseone ND is commissioned. 
\section{Timeline}
\label{sec:timeline}

Design choice decisions for FD3 and FD4 are expected to be made by the DUNE collaboration no later than 2027 and 2028, respectively. The schedule will be driven by the exact details and extent of the enhancements compared to the FD2 design. For example, in the case of a FD2-like module where the only upgrades are optimization of CRPs and the APEX light system, one can envision reaching a final design milestone by 2028 in a technically limited schedule. As for FD1 and FD2, a final design for FD3 and FD4 will be completed after the successful installation of detector demonstrators integrating all system components at full scale, forming "slices" of the full FD system. In this scenario, the earliest start for installation of FD3 (FD4) can be anticipated in 2029 (2030) with completion of installation and argon filling in 2034 (2036). Alternatively, if one were to implement a pixel-based upgrade such as LArPix, Q-Pix, SoLAr, or ARIADNE (top anode plane only), the final design for FD4 would likely not occur until at least 2030. An asymmetric Dual-Phase vertical drift LAr TPC for FD4, with a single drift volume instrumented via an ARIADNE readout plane, would require significant changes to the high voltage system, further pushing out a final design milestone to 2031-32. In the case of the water-based liquid scintillator option, a final design milestone no earlier than 2033 is anticipated.

A detailed schedule for ND-GAr has not yet been developed. A complete conceptual design is anticipated in the late 2020s, with a final design milestone then following in the early 2030s after test beam campaigns with mature hardware prototypes and full slice tests have been successfully completed. Detector installation would commence shortly after, with start of operations expected in the mid-2030s.

\section{Summary}
\label{sec:summary}

DUNE \phaseone is well underway, with key contributions from Europe. In particular, the CERN Neutrino Platform has enabled critical R\&D and validation of the detector technologies for the first two far detectors of DUNE with great success. DUNE relies on continued support from Europe to successfully complete \phaseone and realize \phasetwo. Key requests from DUNE for European Strategy include maintaining ProtoDUNEs at the CERN Neutrino Platform, supporting LBNF infrastructure for DUNE \phasetwo with cryostats for third and fourth far detector modules, supporting European contributions to the \phasetwo near and far detectors and upgraded beamline, and continuing to support scientific and technical leadership in DUNE from Europe. In partnership with Europe, DUNE aims to complete the full scope of DUNE, enabling a multi-decadal program of groundbreaking science with neutrinos.

\newpage



\bibliographystyle{utphys} 
\bibliography{common/references} 

\end{document}